\documentclass[pre,twocolumn, superscriptaddress,nofootinbib]{revtex4}
\usepackage{amsmath,amsfonts,amssymb}
\usepackage[all]{xy}

\begin{document}

\title{Physics of systems with motivation as an interdisciplinary branch of science}
\author{Ihor Lubashevsky}
    \email{ialub@fpl.gpi.ru}
    \affiliation{\mbox{A.M. Prokhorov General Physics Institute, Russian
    Academy of Sciences, Vavilov Str. 38, Moscow 119991, Russia}}
   \affiliation{\mbox{Moscow Technical University of Radioengineering, Electronics, and Automation,
    Vernadsky av. 78, Moscow 119454, Russia}}
\author{Natalia Plawinska}
    \email{n.plavi@mail.ru}
    \affiliation{\mbox{Philology Faculty, Moscow State Pedagogical University, Mal. Pirogovskaya str. 1, Moscow, 119882 Russia}}
\date{\today}
\begin{abstract}
The paper discusses the fundamental characteristics distinguishing the natural and social systems from each other. It considers in detail the basic approaches, prospects, and possibilities of constructing mathematical description for social systems as well as develops the appropriate notions required to do this. The main attention is focused on systems with motion treated as a characteristic example of social systems where the development of mathematical description should demonstrate the crucial ideas of fusing natural and social sciences.
\end{abstract}

%\pacs{02.50.Cw, 02.50.Ey, 02.50.Ga,  05.20.-y, 05.20.Dd, 05.70.Ln}

\maketitle

\section{Introduction}

For many centuries science was much concerned with the laws of the nature. During this time a large variety of physical models based on developed mathematical formalism was created. In the middle of the 19th century there arose a new area of knowledge dealing with objects distinct from the previous ones, the human society and the human behavior. At first, these sciences were only descriptive, then in the course of time some of them, e.g., economics started to use mathematics in their investigations. Several attempts were made to transfer physical concepts and notions to social sciences, which, however, has encountered significant problems. The matter is that the constructs developed for describing natural objects are not always applicable to objects of social nature.

This raises the question as to what notions should be developed and what mathematical formalism has to be constructed in order to describe properly the dynamics and evolution of objects and systems of social nature.

The present paper makes an attempt to find an answer to this question, to formulate the required notions, and to elucidate the way to developing the mathematical formalism for the adequate description of social systems. Since such objects are highly complex and multidimensional we focus out attention on systems with motivation as a relatively simple example exhibiting, nevertheless, many basic properties of social objects discussed in the next section.

\section{Peculiarities of social systems}\label{sec1}

Let us remind some fundamental features distinguishing objects of social nature from ones of the natural world (see, e.g., \cite{1,2}).
\begin{itemize}
  \item (\textit{individuality}) Social systems are made up of elements (individuals, agents, decision makers, etc.) with pronounced individuality in behavior and cognition. By contrast,  for natural systems their elements of one type are identical in properties, for example, all electrons behave in the same way under equal conditions.

  \item(\textit{uncertainty}) The individuality of human beings endows social systems with uncertainty and variability in dynamics. As a particular result, in social systems the regular and random factors are of the same origin and equipollent. The dynamics of natural systems is either deterministic or the regular and random forces arise via different mechanisms.\footnote{This statement does not contradict the uncertainty principle of quantum mechanics, where the wave function obeys the deterministic governing equation. The mechanism responsible for the quantum mechanics uncertainty as well as the stochasticity of statistical systems is up to now is far from being well understood, at least, it is typically assumed to be connected with the openness of physical systems or/and the dynamical chaos.}

  \item(\textit{memory and time constraints}) The laws of social systems can change as the human society evolves. Thereby, first, a specific implementation of these laws should have its onset and a finite lifetime. Second, in studying the regularities of social systems reproducing the initial conditions could be hampered or even impossible. Third, a priory, it is not clear how long the memory of social objects is, in other words, how long time span should separate events in the past from the present instant in order to ignore their effects. Natural systems, by contrast, are characterized by the reproducibility; under the same initial and external conditions either their dynamics or probabilistic characteristics are identical on all the trails. In this meaning, the history of natural systems does not matter.

  \item(\textit{motivation and value factor}) The human behavior is governed by many motives for achieving individual goals as well as obeys the social and cultural norms. There is, typically, a set of possible strategies of behavior among which a decision maker chooses the appropriate one. In doing so he applies to various value factors that reflect his individual preferences and the social and cultural meaning as well. These notions are just inapplicable to natural systems.

  \item(\textit{information deficiency and learning}) The decision-making environment involves many factor that are hidden for decision makers and simultaneously affect substantially the dynamics of social systems. Therefore decision makers seldom have perfect information about the choice alternatives. Thus, they should draw on certain strategies of behavior
      that are based on either own experience or the experience of the society. In the nature such a phenomenon does not exist at all.

  \item(\textit{prediction}) Human beings predict the results of their actions. That is way the system dynamics is affected substantially not only by its history and the current state, but also by its possible future existing in the human mind.

  \item(\textit{breakdown of the explicit means-end relationships}) The dynamics of social systems is affected by a large number of uncontrollable factors, external and internal ones. So if a cause and its effect are separated by a significant time interval it could be difficult to recognize and establish their relationship even in terms of probability. For the disciplines studying natural objects the existence of the direct means-end relationships is one of their cornerstones.
\end{itemize}

Despite these differences, some overlapping between the natural and social sciences has surfaced during the last decades, giving rise to new interdisciplinary branches of science. In particular, methods of statistical physics have turned out to be effective in describing social systems (see a review \cite{3}). The dynamics of complex social systems consisting of many similar elements possesses the self-averaging of the element properties. In other words, to describe their dynamics one can introduce characteristic elements. These elements exhibit a regular behavior and take into account the human individuality in terms of random factors whose probabilistic properties are identical for all the elements of one type. Such an approach enables us to overcome the main hurdle for the interpenetration of social and natural sciences caused by the human individuality and puts forward new objects like the characteristic individual, the typical decision maker, etc. \cite{4} These notions are the generalized images that characterize the common and reproducible properties of human beings. They form the basis of the appropriate language for tackling social systems using the methods developed in physics, mathematics and other disciplines of natural sciences.

It is worthwhile to underline that there are conceptual differences between the characteristic elements of social systems and ones of natural systems (cf. \cite{3}). In particular, the notions of probability form the basic language in modeling social systems. For comparison: the elements of systems of classical physics are described by deterministic laws on the ``microscopic'' level and only on the ``mesoscopic'' level their dynamics becomes stochastic and the probabilistic language becomes appropriate for tackling their complex behavior. As another instance of such a differences we note the following. For example, electrons and atoms are relatively simple objects with well known properties. The macroscopic complex phenomena observed in their ensembles stem from the interaction of a large number of `simple' elements rather than are due to their complexity. For social systems the situation is different. Human individuals are entities with complex behavior being already the outcome of many internal psychological and physiological processes. So a complex behavior of social systems can be caused by two factors. The first one is again the interaction of many individuals, the second one is related to the individual complexity of human beings whose behavior involves variety of dimensions. So in tackling a specific social phenomenon it is necessary beforehand to single out which dimensions should be taken into account. Moreover, if several aspects of human being play key role in the analyzed phenomenon there is a possibility that its dynamics will be governed by various regularities with distinctive features.

\section{Decision-making process}\label{sec3}

Let us discuss in more detail the decision-making process, which is one of basic factors governing the dynamics of any social system. The classical decision theory is based on the notion of the preference relation and the utility function quantifying this relation (see, e.g., \cite{5}). The concept of the perfect rationality assumes the human choice or decision to be determined by the most preferable result meeting the maximum of the utility function. The classical theory of decision under uncertainty also deals with some utility function aggregating in itself the realization of various environmental conditions in a probabilistic way. However, such an approach encounters obstacles caused by the fundamental properties of human beings.

First, any decision or choice should be made during a certain finite time interval. On one hand, such time constraints together with the state's multitude of the decision-making environment prevent an individual from identifying adequately the current conditions or estimating the probability of their realizations (\emph{uncertainty of information about the system states}). On the other hand, it hampers the proper ordering of the actions and aims according to their preference and priority. Various states of the decision-making environment or various actions and aims of individuals that are similar in value can be indistinguishable for human beings in making decisions (\emph{uncertainty of event value}).

Second, making a decision is evaluated, at least, by two type factors. One of them is the value of the result, the other is related to the cost of efforts required to do this, including gathering and processing information under the time constraints. So both these aspects affect significantly the human behavior. Besides, the time endowment of making a decision is likely not to be an independent characteristics but to stem from a certain compromise between the two factors. The feasibility of introducing the preference relation with respect to the second factor seems to be doubtful, at least, it is an open problem. Its solution is complicated by the fact that the information about the decision-making environment can be accumulated, thereby, the history of the system dynamics becomes essential (\emph{memory effects}). The latter feature, in particular, poses a problem about the feasibility of initial conditions for social systems.

Third, there could be a long time span between the moments of making a decision and getting its results with a large number of intermediate factors. These factors affect the system dynamics and are hidden for decision makers. As mentioned above, it breaks down the explicit means-end relationship.

These limitations in human cognition induced the development of the concepts of bounded rationality \cite{6,7} and limited cognition \cite{8} (see also reviews \cite{9,10,11}). Under these limitations the concept of perfect rationality is inapplicable to describing the decision-making process, at least, directly. So there should be another mechanism governing the decision-making. It has been proposes \cite{11} that the decision-making process in the social system dynamics is based on possible strategies of behavior rather than the choice of final goals. Such a strategy is a certain sequence of local actions, i.e. a collection of steps of achieving subsequent intermediate aims. These strategies are the results of trial-and-error process and evolve during the adaptation of individuals to the decision-making environment under uncertainty of information about the states of the social system and its dynamics. Following \cite{11} we will call these strategies heuristics.

These heuristics aggregate and accumulate the information about the previous actions, successful and failed ones. That is way the history of a social system impacts on it dynamics. There are at least two distinct ways of the heuristics formation.

The first one is the individual learning, i.e. the process of gaining the knowledge about the successful rules of behavior via the personal experience or the experience of individuals directly related to a given one. In particular, the idea that the individual learning plays the leading role in the heuristics formation has been developed in \cite{12,13} (see also references therein).

The second way matches the cooperative interaction of many individuals forming large units of human society. It is implemented via the formation of the social norms and cultural values aggregating all the fragments of information about the human society for a rather long time interval. The human societies possess own mechanisms governing the social norms and keeping up the social order (see, e.g., \cite{11} and references therein). It should be pointed out that this idea about the social norms is not new; it goes back to works by Veblen \cite{14} and Pigou \cite{15}.

There are at least two types of models for mechanisms via which the social norms and cultural values arise and evolve. One of them is based on emulating the behavior of the most successful persons, i.e. the social \emph{interdependence via significant others} \cite{16} (for a brief review of relative models see also \cite{11}). The other type models, \emph{interdependence via reference groups} \cite{17}, go beyond the individualistic level of social interdependence. They relate the social and cultural proclivities of human behavior to some large groups or their typical representatives that have high social rank. In the latter case the heuristics can be accumulated within such reference groups, which enables us to regard them as some social capital \cite{11}.

The aforementioned features allow us to declare that a new own mathematical formalism is necessary in order to describe social systems. In particular, the appropriate notions and concepts should be concerned with the bounded capacity of human cognition affecting the decision-making process. The same statement holds also for systems of other types where, however, the human factor plays an essential role.

\section{Systems with motivation}

In what follows, we will consider a specific type of social systems that can be categorized as systems with motivation. The systems with motivation are actually a relatively simple example of social objects that, nevertheless, exhibits all the general properties discussed above. The behavior of their elements is governed by the decision-making under bounded rationality and the leading role of local motives, being the reason for the used name. Another essential feature of such systems is the fact that they are made up of many elements, which in turn can be divided into large groups by similarity. Therefore the self-averaging property should hold, enabling us to introduce the corresponding characteristic elements.

In the present section we make an attempt to construct the notions and concepts required for describing systems with motivation and to discuss the appropriate mathematical formalism, including the general form of the governing equations. In some sense this mathematical formalism can be treated as a detailed explanation of the essence of systems with motivation. Their particular examples will be mentioned in the text below.

First of all, it is necessary to determine the phase space of a given system in order to describe its dynamics. Due to the active cognitive behavior of its elements the phase space, $\{w\}$,  comprises variables of two types, objective and subjective ones, $\{w\}=\{q,h\}$. Let us discuss them individually.

\textit{Objective phase space:}
There is assumed to be a collection of variables $\{q\}_\alpha$, discrete or continuous ones, that completely characterize the possible states of a given element $\alpha$ from the standpoint of the other elements. The information about the current state of the element $\alpha$ is necessary for them to make the appropriate decisions in governing their own states.  We have used the ``objective'' term in order to underline that the characteristics $\{q\}_\alpha$ of the element $\alpha$ are detectable for the other elements. They are not related to the intentions, plans, wishes of the element $\alpha$ which are hidden for external observers. It should be noted that the variables $\{q\}_\alpha$ are accessible for external observers only in principle. As noted above in a social system getting information about the states of its elements can be hampered. The quantities $\{q\}_\alpha$ will be referred to as the objective phase variables ascribed to the element $\alpha$ and their combination for all the elements, $\{q\}$, makes up the objective phase space of the given system. For example, the spatial coordinates of pedestrians, the direction of motion, and may be their velocities form the objective phase space of the pedestrian ensemble, the coordinates and the velocities of vehicles on a highway make up the objective phase space of traffic flow. The set of personal opinions makes the objective phase space of voting process, cultural features with preferences ascribed to every individual can be regarded as the objective phase space of the cultural dynamics. The production and comprehensive matrices characterizing the frequency of using and associating words to the corresponding objects by individuals can be considered in the same way in describing the evolution of languages (see \cite{3} and references therein).

The phase space of natural objects may comprise only measurable (objective) quantities. Therefore, using the notions and concepts inherited from physics the majority of mathematical models proposed for social systems assume their dynamics to be completely determined by the corresponding objective phase space (see, e.g., reviews \cite{3,Helbing}). These models, however, are applicable to social systems as a rough approximation only because of takeing into account just a few of their basic features.

In turn, among the objective phase variables a special group of controllable quantities should be singled out. Elements of social systems try to control their own states, which influences the system dynamics. This self-control is determined, on one hand, by individual motives, desires, wishes, goals, plans etc. and, on the other hand, by the social order and the rules of behavior. It is implemented via maintaining or changing the objective phase variables or a certain group of them enabling this action directly. The subscript $c$ will be added to the corresponding quantities $\{q_c\}$ to underline the given feature. Time variations of the remaining quantities are determined by these controllable variables and, may be, some natural regularities. For example, in traffic flow a driver can change directly only the velocity of his car. Therefore the velocities of cars are the controllable variables, whereas the coordinates of their position on highways are not so.

\textit{Subjective phase space:}
Let us introduce new quantities $\{h\}$ to denote time variations in the controllable phase variables, $\{h:=\delta_t q_c\}$. Here the symbol $\delta_t$ stands for the rate of time change in the corresponding quantity if it is continuous one or, otherwise, describes step-like jumps between its possible values. The collection of quantities $\{h\}_\alpha$ determined directly by a given element $\alpha$ will be referred to as the subjective (hidden) phase variables of this element and the combination of all these quantities will be called the subjective phase space of the given system.

The quantities $\{h\}_\alpha$ characterize certain internal processes in making decisions by the element $\alpha$. So they are accessible only for the element $\alpha$ and hidden for others. For every element $\alpha$ its subjective phase variables $\{h\}_\alpha$ are valuable in their own right. This is due to the fact that internal processes accompanying the decision-making themselves take effort in order, for example, to get a decision of changing the current state of the element $\alpha$. In addition, time variations in the quantities $\{q_c\}$ can affect this element in some physical way. Therefore, in making decision the preferences are determined directly by the objective and subjective variables simultaneously. That is why the phase space of systems with motivation is made up of both the types of the phase variables, $\{w\} = \{q,h\}$.

\textit{The general form of governing equations:}
The decision making process governs time dependence of the controllable objective variables, which in turn determines the system dynamics. Symbolically we write this in terms of time increment in the phase variables
\begin{equation}\label{1}
\begin{split}\xymatrix{
    \text{decision-making}\ar@{:>}[d] & \{w\} = \{q,h\}\ar@{.>}[d]
\\
     \{h=\delta_t q_c\}\ar@{:>}[r] &\{\delta_t w\}\,.
}
\end{split}
\end{equation}
However, in the general case expression~\eqref{1} does not represent any function directly relating the phase variables $\{w\}$ and their time variations. The phase space even extended in such a way does not enable one to specify the dynamics of a social system because of the bounded capacity of human cognition. The whole history or, at least, its long fragment determines the dynamics of a given system. In mathematical terms we have to deal with the whole trajectory $\{w[t']\}_{t'<t}$ of the system motion in the extended phase space $\{w\}$ at the previous moments of time, $t'<t$, in order to find the increments $\{\delta_t w\}$ in the phase variables at the current moment of time $t$. Therefore the governing equation of the system dynamics should be represented as
\begin{equation}\label{2}
\xymatrix{
    \{w[t']\}_{t'<t}\quad \ar@{|->}[r] &\quad \{\delta_t w\}\,.
}
\end{equation}
Here the square brackets at the symbol $w$ stand for the function $w$ of the argument $t'$ rather than its value taken at $t'$.

According to diagram~\eqref{2}, the dynamics of systems with motivation does not belong generally to the class of initial value problems as it is the case for natural objects. It is likely that the closer a \emph{given} event in the past to the current moment $t$, the larger its contribution to the system dynamics at the present time. So functional~\eqref{2} seems to be some integral over time with a kernel decreasing as the analyzed point in the past goes away from the current moment of time.

The structure of functional \eqref{2} has to take into account several features, in particular, choosing strategies of behavior for achieving the desirable aims, the bounded perception of human beings, and the role of social and cultural norms. Let us consider them individually detailing diagram~\eqref{2} step by step.

\textit{Heuristics choice:}
As noted in the previous section, the decision-making process is reduced to the choice of local heuristics because of the bounded capacity of human cognition and the variety of factors uncontrollable and hidden for the elements. These heuristics, i.e., the local strategies of the element behavior are sequences of actions focused on achieving local aims. Since in social systems the explicit means-end relationships can be broken the specific actions of elements are evaluated by local motives rather than intentions of getting final goals. The latter goals can only single out some rather general class of the element actions. Moreover the final goals are typically stated in a rather general form without particular details.

In order to describe the heuristics choice we introduce an imaginary phase space $\{\varpi\}_\alpha$ in addition to the real one $\{w\}$ which is ascribed individually to each element $\alpha$. So we have actually introduced the set of spaces existing actually in the human mind. The imaginary phase spaces enable us to specify a hypothetical dynamics of the system in the near feature that can be expected by its elements based on the available information. Every imaginary phase space
\begin{equation}\label{ips}
   \{\varpi\}_\alpha = \{\boldsymbol{\theta},\eta\}_{\alpha}
\end{equation}
comprises the objective variables $\{\boldsymbol{\theta}\}_\alpha$ of all the elements and the subjective variables $\{\eta\}_\alpha$ of the given element $\alpha$. These quantities specify the hypothetical states of the elements in the ``mind'' of the element $\alpha$. Therefore the symbol $\boldsymbol{\theta}_{\alpha':\alpha}$ is written in bold to underline its dependence on two indices, meaning the phase variables of the element $\alpha'$ in the ``mind'' of the element $\alpha$. Collection~\eqref{ips} does not contain the subjective phase variables of the other elements because they are hidden for the given element $\alpha$.

In these terms a possible strategy of behavior of the element $\alpha$ is represented as a certain time dependence $\{\eta[t'']\}^{t''>t}_\alpha$ of its subjective phase variables in the near future. The hypothetical time dependence $\{\boldsymbol{\theta}[t'']\}^{t''>t}_{\alpha:\alpha}$ of its objective variables is determined by the given strategy of behavior. The hypothetical time dependence $\{\boldsymbol{\theta}[t'']\}^{t''>t}_{\alpha':\alpha}$ of the objective variables ascribed to another element $\alpha'\neq\alpha$ is constructed in the ``mind'' of the element $\alpha$ based on the available information. We also will use the notation $\{\varpi[t'']\}^{t''>t}_\alpha$ to denote this strategy as the hypothetical motion of the system in the space $\{\varpi\}_\alpha$. The symbol $\{\varpi[t'']\}^{t''>t}$ without the element index stands for the heuristics as whole.

The elements are assumed to evaluate and choose the desired strategies of behavior $\{\varpi_\text{op}[t'']\}^{t''>t}$ in some optimal way, which determines the system dynamics. It should be noted that in this choice every element $\alpha$ evaluates possible strategies of its own behavior $\{\varpi_\text{op}[t'']\}^{t''>t}_\alpha$ only, the behavior of the other elements is regarded by it as given beforehand or predictable with some probability. These features of the heuristics choice enable us to represent symbolic expression~\eqref{2} as
\begin{equation}\label{3}
\begin{split}
\xymatrix{
   \{w[t'],\varpi[t'']\}^{t''>t}_{t'<t}
    \ar@{=>}[rr] ^-{\text{individual choice}}_-{\text{of system elements}}
    \ar[dr]
    && \ar@{=>}[dl]^{\quad\{h[t'']\}^{t''>t}} \{\varpi_\text{op}[t'']\}^{t''>t}\\
    & \{\delta_t w\}
}
\end{split}
\end{equation}
It should be pointed out that in choosing the heuristics the elements can predict the system dynamics extrapolating the time variations of the phase variables in some simple way, for example, fixing them or supposing the linear time dependence to hold in the near future.

\textit{Perfect rationality:}
In order to find a measure for quantifying the heuristics let us consider a certain limit case called the perfect rationality. It comes into being when, first, analyzed situations are repeated many times, with the environment conditions being the same. Thereby the time restrictions affecting the decision-making process are removed and the complete information about the system becomes accessible. Second, the elements are able to correct their states continuously.

Under such conditions the individual choice of the optimal heuristics $\{\eta_\text{op}[t'']\}^{t''>t}_\alpha$ by a given element $\alpha$ is reduced to finding the maximum of a certain preference functional
\begin{equation}\label{4}
     \mathcal{U}_\alpha := \mathcal{U}\Big\{\{\varpi[t'']\}^{t''>t}_\alpha, \{w[t']\}_{t'<t} \Big\}
\end{equation}
with respect to its own strategy of behavior $\{\eta[t'']\}^{t''>t}_\alpha$. In the general case functional~\eqref{4} depends also on the element type, which, however, is not labeled directly for the sake of simplicity. Functional~\eqref{4} quantifies the preferences of the element $\alpha$ in the choice of its own heuristics, provided the system history and the behavior of the other elements are known beforehand. In other words, within the frameworks of the perfect rationality the optimal strategy of behavior $\{\eta_\text{op}[t'']\}^{t''>t}_\alpha$ is specified by the expression
\begin{equation}\label{5}
  \{\eta_\text{op}[t'']\}^{t''>t}_{\alpha}
  \quad\Longleftarrow\quad
  \max_{\{\eta[t'']\}^{t''>t}_\alpha}
    \mathcal{U}_\alpha\,.
\end{equation}
By way of example, we note that in the limit of perfect rationality schema~\eqref{3} for traffic flow gives rise to Newtonian type models \cite{18}. It is the case where the concept of social forces \cite{Helbing} holds. If the driver behavior is not perfect then the description of traffic dynamics goes beyond the notions of Newtonian mechanics \cite{20}.

Expression~\eqref{5} specifies the optimal heuristics $\{\eta_\text{op}[t'']\}^{t''>t}$ as certain trajectories. Therefore, if the elements with perfect rationality choose these optimal strategies of behavior at the current moment of time and follow them, then further correction of the system motion will be not necessary. The latter is the essence of the Nash equilibrium.

We assume that the analyzed system admitting the limit case of perfect rationality possesses also a special point $\mathfrak{Q}$ (or a set of points with equal values of the controllable variables, $\{q_c = \text{const}\}$) in the objective phase space. This point united with the origin $\{\eta = 0\}$ of the subjective phase space matches the steady-state dynamics of the given system under stationary external conditions in the limit of perfect rationality. It means that if the system is initially located at the given point, it will not leave this point further. For example, traffic flow where all the cars move with the same speed and at some optimal headway distance matches this situation. If the system during its motion governed by expression~\eqref{5} tends to the point $\mathfrak{Q}$ or the corresponding set of points, it will be referred to as an attractor of rational dynamics.

\textit{Bounded rationality:}
As discussed previously, the time constraints together with the bounded capacity of human cognition endow the choice of heuristics and, thereby, the system dynamics with random properties. If two strategies of behavior are rather close to each other in value then it can be tough to order them by preference and to choose one in a rational way. We apply the notion of perception threshold $\Theta$ to tackle this problem. The perception threshold as well as the preference functional depends generally on the type of elements, which again is not labeled directly to simplify the notations.

Let us make use of the preference functional~\eqref{4}. Two strategies of behavior $\{\eta_1[t'']\}^{t''>t}_\alpha$ and $\{\eta_2[t'']\}^{t''>t}_\alpha$ are considered to be equivalent, with the other environment conditions being the same, if the corresponding magnitudes of the preference functional meet the inequality $\left|\mathcal{U}_{1,\alpha}-\mathcal{U}_{2,\alpha}\right| \lesssim \Theta$. It is the point where the form of the preference functional~\eqref{4} becomes determined. The matter is that any increasing function applied to the preference functional gives rise to a new preference functional describing the same set of the optimal heuristics. Introducing the perception threshold we actually fix its form.

When a currently chosen strategy of behavior $\{\eta[t'']\}^{t''>t}_\alpha$ is close to the optimal one in the given sense, the element $\alpha$ has no motives to change it. Roughly speaking, if it is not clear what to do, to change nothing is quite adequate. If the difference in the magnitudes of the preference functional~\eqref{4} for the two strategies becomes remarkable in comparison with the perception threshold, the element $\alpha$ recognizes the necessity of correcting its current state. Exactly this choice of new more proper heuristics is the point where the random factors enter the system directly. Indeed, since all the strategies of behavior that are close to the optimal heuristics in terms of the perception threshold are regarded as equivalent then the choice of some of them is a random event. The time moment when this choice arises is also a random quantity.

It should be pointed out that the perception threshold $\Theta$ characterizes probabilistic properties of the element behavior rather than the step-like dynamics. Namely, let us consider two heuristics, the strategy of behavior $\{\eta_c[t'']\}^{t''>t}_\alpha$ that is followed by the element $\alpha$ at the current moment of time $t$ and the optimal one $\{\eta_\text{op}[t'']\}^{t''>t}_\alpha$ which is actually hidden for it. When the difference in the corresponding magnitudes of the preference functional~\eqref{4} becomes equal to the threshold, $|\mathcal{U}_{c,\alpha} -\mathcal{U}_{\text{op},\alpha}| = \Theta$, the probability of correcting the current state by the element $\alpha$ just attains its maximum rather than exhibits a very sharp pike. In the cases $|\mathcal{U}_{c,\alpha} -\mathcal{U}_{\text{op},\alpha}| \ll \Theta$ the element cannot recognize the fact of the system deviating from the optimal dynamics. States matching the opposite inequality $|\mathcal{U}_{c,\alpha} -\mathcal{U}_{\text{op},\alpha}| \gg \Theta$ cannot be reached because the element would respond earlier. In particular, according to empirical data for traffic flow such events of correcting the car motion are distributed rather widely near the corresponding threshold \cite{24}.

\textit{Action points:}
Let us introduce the notion of action points in order to describe the dynamics of systems with motivation. An action point is an event of changing the current strategy of behavior by some element in correcting its state. Every action point is associated with this strategy and the time moment of changing it. When the dynamics of a given system is optimal within the human perception characterized by the threshold $\Theta$ its elements do not correct their heuristics. At these moments of time the system motion is not governed by the elements and proceeds according to natural regularities affecting the system. When the system motion deviates from the optimal one substantially the elements recognize this fact and correct their individual strategies of behavior. In doing so an element selects some new strategy of behavior in a neighborhood of the optimal heuristics whose thickness quantified by the preference functional~\eqref{4} is less than or of the order of the perception threshold. Then the given element follows the selected heuristics until it recognize the necessity of its state correction again. We note that the notion of action points for the car-following process was introduced for the first time in \cite{23} to denote the moments of time when drivers correct the motion of their vehicles.

In these terms the dynamics of a system with motivation can be represented as a sequence of action points, i.e. jumps between various strategies of the element behavior. The particular strategies of behavior joined by these jump-like transitions and their time moments are random quantities. These random transitions are the cause of the stochasticity in the dynamics of systems with motivation. Between the action points the system dynamics is not governed by its elements at all and is regular or affected by random factors of natural origin. Symbolically this feature is represented by the following diagram generalizing the previous one~\eqref{3}

\begin{equation}\label{6}
\begin{split}\xymatrix{
   \{w[t'],\varpi[t'']\}^{t''>t}_{t'<t}
    \ar@{=>}[rr] ^-{\text{individual choice}}_-{\text{of system elements}}
    \ar@{->}[d]
    && \ar@{=>}[dl]^{\text{\quad action points}} \{\varpi_\text{op}[t'']\}^{t''>t}\\
    \{\delta_t w\} & \{\delta_t h\}\ar@{=>}[l] &
}
\end{split}
\end{equation}
If such a system possesses an attractor of rational dynamics it can exhibit a new type of cooperative phenomena. Indeed, by definition,  the point $\mathfrak{Q}$ in the objective phase space matches the steady-state dynamics of a system with perfect rationality. Then the elements with bounded rationality will regard the system motion in the vicinity of this attractor also optimal. To discuss the given feature in more detail let us introduce the notion of dynamical traps.

\textit{Dynamical traps:}
Using the preference functional~\eqref{4} we construct a certain neighborhood of the set $\mathfrak{Q}\bigotimes\{h=0\}$ in the space of heuristics $\{\varpi[t'']\}^{t''>t}$ of thickness $\Theta$ and then project it onto the objective phase space $\{w\}$. In this way we obtain a certain neighborhood $\mathfrak{D_Q}$ of the set $\mathfrak{Q}$ called the dynamical trap region. When the system with bounded rationality enters this region its elements consider the system dynamics optimal and the correction of their state unnecessary. Since the system dynamics in the region $\mathfrak{D_Q}$ is really close to the optimal one, a time span between two action points could rather prolonged in comparison with that of the system dynamics far from $\mathfrak{D_Q}$. In other words such fragments of system motion inside the dynamical trap region can be regarded as long-lived states \cite{DT1,DT2}. Their origin is due to the stagnation of the element active behavior for a relatively long time. Therefore dynamical traps are able to induce nonequilibrium phase transitions of a new type that should be widely met in social systems \cite{DT1,DT2,DT3} rather than in natural systems. We note that the dynamical traps for Hamiltonian systems was introduced in \cite{Zas1,Zas2} (see also a review \cite{ZasRev}) and for systems with nonlinear oscillations it was done in \cite{Gaf}. 

\textit{Individual learning and formation of social and cultural norms:} 
Because of the bounded capacity of human cognition, gaining the knowledge about the proper strategies of behavior is crucial. As noted in Sec.~\ref{sec3} there are two channels of accumulating and aggregating such information. One is the individual learning of the elements based on own experience or local interaction with the other elements. In some sense it is a typical mechanism of cooperative phenomena widely met in natural systems and caused by local or quasi-local interaction of their particles. It seems to be possible to describe the individual learning process using the introduced phase space and perception thresholds. No additional variables are necessary to do this. Indeed, let us ascribe an individual perception threshold $\Theta_\alpha$ to every element $\alpha$. Then the individual learning is represented as the evolution of the perception thresholds $\{\Theta_\alpha\}$ caused by some interaction of the elements. Symbolically it takes the form
\begin{equation}\label{77}
    \{w[t']\}_{t'<t} \Longrightarrow \{\Theta_\alpha(t)\}\,.
\end{equation}
In these terms the individual learning is reduced to the time decrease of the perception thresholds $\{\Theta_\alpha\}$ due to the accumulation and aggregation of information about the system properties.

The second channel is related to a unique collective interaction of all the elements in a social system in addition to their individual interrelations of various types. It arises via the formation of social and cultural norms of behavior. These norms affect directly the heuristics and their preference, and involve all the members of a social system or their large groups independently of their relationships and distance in space and time. The social and cultural norms aggregate the information about the properties and features of a social system during a long time interval and make up the basis for finding \emph{general} rules of successful strategies of behavior. So in order to describe the effect of the social and cultural norms on the system dynamics some additional variables, the space of cultural features $\{\chi\}$, should be introduced. We presume that the cultural features cannot be ascribed to individual persons in any way, they have their own carriers, e.g., books, newspapers, magazines, movies, broadcasts, and other types of mass media. Symbolically the formation of social and cultural norms can be written as
\begin{equation}\label{7}
    \{w[t']\}_{t'<t} \Longrightarrow \{\chi(t)\}\,,
\end{equation}
where, as it is rather natural to assume, the father a \emph{given} event in the past, the weaker its influence on the present. In order to include the effect of these norms on the social system dynamics we generalize diagram~\eqref{6} as follows
\begin{equation}\label{8}
\begin{split}\xymatrix{
   \{w[t']\}_{t'<t} \ar@{=>}[r] \ar@{~>}[rd]_{\text{individual learning\qquad}}
      & \{\chi(t)\} \ar@{~>}[d]
\\
   \{w[t'],\varpi[t'']\}^{t''>t}_{t'<t}
    \ar@{=>}[rr] ^-{\text{individual choice}}_-{\text{of system elements}}
    \ar@{->}[d]
    &{\protect\phantom{\Big(sssssss\Big)}}& \ar@{=>}[dl]^{\text{\quad action points}} \{\varpi_\text{op}[t'']\}^{t''>t}
\\
    \{\delta_t w\} & \{\delta_t h\}\ar@{=>}[l] &
}
\end{split}
\end{equation}
which is the final diagram presenting the essence of the mathematical components and notions describing the systems with motivation.

\section{Conclusion}

The paper has considered systems with motivation as a typical example of social systems, where human factor plays a key role. First, the elements of such a system are characterized by motivated behavior and the decision-making process governs the system dynamics. Second, all the elements can be divided into large groups by their properties and similarity. Therefore the self-averaging holds in these system, enabling us to introduce the notion of the characteristic elements. Their regular properties describe the common characteristics of the element behavior, whereas the random ones take into account the individuality of the elements as well as the unpredictable features in their behavior.

Because of the bounded capacity of human cognition the decision-making process is reduced to the choice of heuristics, the strategies of the element behavior focused on achieving a sequence of local aims rather than a final goal. According to the general properties of social systems the decision-making processes under consideration is governed by the following factors:
\begin{itemize}
  \item local motives and stimuli because the explicit long-time means-end relationship is usually broken by a large number of intermediate uncontrollable effects,
  \item uncertainty and deficiency of information about the current state of the system and the decision-making environments,
  \item time constraints and bounded human ability to order aims according to their preference,
  \item the crucial role of accumulating information about the system properties and the proper behavior under various conditions based on the history of the system dynamics, which is implemented via individual learning of elements as well as aggregation of the information in the common social and cultural norms.
\end{itemize}

Since these features affect the system dynamics substantially all of them should be reflected in the corresponding mathematical notions in describing systems with motivation. In particular, the appropriate mathematical language has to be concerned with the following.
\begin{itemize}
  \item A system with motivation is characterized by the extended phase space including the objective variables as well as the subjective ones. The objective variables describe the states of the system elements and their values are accessible, in principle, for all the elements. The subjective variables characterize individual motives and plans of the elements and their values are accessible only for the own elements. More rigorously, every subjective variable specifies time variations in one of the objective variables changing which the element affects the system dynamics.

  \item The motion of the system at a given moment of time is determined by the trajectory of its motion at the previous moments of time as a whole. In other words, the system dynamics is affected directly by the history of the system rather than the particular point of the phase space at which the system is currently located. Thereby the dynamics of systems with motivation is not described by the initial value problem.

  \item The dynamics of a system with motivation is governed by the individual decision-making process of its elements. However, the explicit means-end relationships can be broken by a large number of intermediate uncontrollable factors. Therefore the specific actions of the elements are not determined directly by the final goals distant in time. These goals can only single out some rather general class of actions. Moreover the final goals are typically stated in a general form without particular details. As a result, governing the system dynamics is reduced to choosing between possible strategies of the element behavior focused on achieving a sequence of local aims rather than on getting the final goals. The previous trial-and-error experience of the elements affects significantly the formation and evolution of these strategies of behavior, so they are also called heuristics. In mathematical terms these heuristics are represented as trajectories of possible system motion in the near future. Therefore the preference relation connected with the given decision-making process should be addressed to the these trajectories.

  \item Stimuli and motives for actions of individual elements determine the choice of such heuristics that are optimal or considered to be optimal by the elements. This choice is affected by the prognosis of the changes in the system states in the near future. Thereby the imaginary future influences the real present, and vice versa.

  \item The systems with motivation are assumed to conform to the principles of the perfect rationality in the limit case when the analyzed situations are repeated many times, with the environment conditions being equal. Under these conditions the time restrictions and the uncertainty of information about the system state are is removed. Thereby the heuristics can be rigorously ordered by some preference relation, enabling a preference functional to be introduced.

  \item The time restrictions, the deficiency of information about the system states, and the bounded capacity of human cognition give rise to the uncertainty in the preference and choice of these heuristics. So if two strategies of behavior turn out to be quite similar in quality and value then they are treated as equivalent and their choice is equiprobable. Quantitatively this effect called the bounded rationality is described by the threshold of element perception. Namely, if the difference between the corresponding values of the preference functional is less then or order of the perception threshold than these strategies are regarded by the elements as equivalent. It should be pointed out that exactly the perception threshold removes the ambiguity in the specific form of the preference functional.

  \item Due to the bounded rationality, the dynamics of systems with motivation is endowed with stochastic properties via the action point mechanism. Namely, when the system motion deviates substantially from the optimal one the elements notice that their states are not optimal. An element can do this when the value of the preference functional for its current state differs from that of the optimal one by the perception threshold. This difference causes the element to correct its current strategy of behavior choosing new one near the optimal heuristics. The choice of this new heuristics as well as the moment of this action is a random event. As a result, the system dynamics looks like a sequence of random jumps discrete in time between the possible strategies of behavior, i.e. the sequence of action points. Between these events the system moves without direct control by elements. We would like to underline that the perception threshold specifies the preference functional variation when the correction of the element states becomes most probable rather than the step-like behavior of the elements.

  \item If in the objective phase space there is a point matching the steady-state dynamics in the approximation of the perfect rationality, then for this system it is possible to introduce the notion of the dynamical trap region being a certain neighborhood of the given point. Entering the dynamical trap region the system spends inside it a relatively long time before leaving it because of stagnation of the element behavior. The dynamical traps and local interaction of elements can give rise to a multitudes of long lived states and nonequilibrium phase transition of a new type caused by the stagnation of the system motion in the dynamical trap region. These phase transitions should be typical for social systems rather than natural ones.

  \item Because of the bounded cognition of human beings, the knowledge about the proper heuristics are gained and accumulated based on the history of the system dynamics for a relatively long period of time. This process is implemented via two channels. The first one is individual learning of elements and can be described by evolution of the individual perception thresholds caused by local interaction of the elements with one another. The second channel is related to the common social and cultural norms. They are responsible for the interaction involving all the elements (individuals) independent of their relationships and distance in space and time. Therefore the social and cultural norms cannot be ascribed to the individual elements, and there should be a certain special medium carrying these norms.

\end{itemize}

Summarizing the aforesaid we claim that developing physics of systems with motivation it could be possible to fuse the social and natural sciences within common notions and approaches. It is likely that the ``microscopic'' language appropriate for this purpose should deal with the trajectories of system dynamics as the elementary objects.

\acknowledgments

The work was partially supported by DFG Grant MA 1508/8-1 and RFBR Grants 06-01-04005 and 09-01-00736.

\end{document}